\documentclass{pasa}%

\usepackage{graphicx}
\usepackage{mathtools}

\def\fig{Fig.}

\def\Fig{Figure}

\def\sect{Sect.}

\def\tab{table}
\def\tabs{tables}
\def\Tab{Table}

\def\eqn{equation}
\def\eqns{equations}
\def\Eqn{Equation}

\title[A Measurement of Source Noise]{A Measurement of Source Noise at Low Frequency: Implications for Modern Interferometers}

\author[J. S. Morgan et al.]{J .S. Morgan,$^{1}$ R. Ekers,$^{1,2}$
\affil{$^{1}$International Center for Radio Astronomy Research, Curtin University, GPO Box U1987, Perth, WA 6845, Australia}
\affil{$^{2}$CSIRO Astronomy and Space Science (CASS), P.O. Box 76, Epping, NSW 1710, Australia}
}

\jid{PASA}
\doi{10.1017/pas.\the\year.xxx}
\jyear{\the\year}

\usepackage{aas_macros}
\usepackage{hyperref} 
\hypersetup{colorlinks,citecolor=blue,linkcolor=blue,urlcolor=blue}

\begin{document}

\begin{frontmatter}
\maketitle

\begin{abstract}
We report on the detection of source noise in the time domain at 162\,MHz with the Murchison Widefield Array.
During the observation the flux of our target source Virgo A (M87) contributes only $\sim$1\% to the total power detected by any single antenna, thus this source noise detection is made in an intermediate regime, where the source flux detected by the entire array is comparable with the noise from a single antenna.
The magnitude of source noise detected is precisely in line with predictions.
We consider the implications of source noise in this moderately strong regime on observations with current and future instruments.
\end{abstract}

\end{frontmatter}
\section{INTRODUCTION}
Source noise (also known as self noise, wave noise or Hanbury Brown Twiss noise) arises from the fact that most sources studied in radio astronomy are themselves intrinsically noise-like: i.e. stochastic, ergodic, Gaussian random noise \citep[][1.2]{1999ASPC..180..671R,2017isra.book.....T}. 
Since these natural sources are typically very weak relative to other sources of noise (referred to here and elsewhere as ``system noise'') this contribution can almost always be neglected.
However, as telescopes are built with increasing sensitivity we are more likely to approach or even reach the strong source limit.
In this limit (i.e. where the signal strength dominates over other types of noise such that generated in the amplifier) the noise becomes proportional to the signal strength and so the signal to noise ratio becomes independent of the sensitivity of the instrument: the only way to increase the signal to noise is to obtain further independent samples by increasing either the bandwidth or observing time. 
Larger collecting area or more antennas no longer help.

For a single dish, source noise is straightforwardly understood as a contribution to the total noise when the source is in the field of view.
For interferometry the interpretation is a little less straightforward and source noise is better understood as distinct from other noise with distinctive properties.
This is because unlike other types of noise, each interferometer element receives the \emph{same} source noise from an unresolved source (i.e. the source noise is correlated between antennas).
It is this correlation which is exploited in an intensity interferometer \citep{1954PMag...45..663B}.

This has two important consequences: first, just like the single-dish case, source noise is only seen at the location of the source in either an aperture synthesis image or a phased-array beam.
More accurately, in a synthesis image, the noise is distributed with the shape of the instantaneous Point Spread Function (PSF).
Secondly, for an interferometer with a large number of elements, source noise may contribute significantly to the image noise at the location of the source even if it is dwarfed by the system noise of a single interferometer element, since unlike noise which is uncorrelated between elements, the noise cannot be reduced by averaging over baselines\footnote{Note that coherence across baselines, a concept fundamental to interferometry, should not be confused with temporal coherence. One does not imply the other}.

This intermediate regime, where the source is weaker than the noise, but is not weak enough to be negligible, is the subject of this paper.
We present a clear detection of source noise (arguably the first in the intermediate regime) in data from the Murchison Widefield Array \citep[MWA; a 128-element low-frequency interferometer;][]{2013PASA...30....7T}.
We then examine the scenarios in which this may have a negative impact.

The paper is organised as follows:
in \sect~\ref{sec:review} we review the literature on source noise in interferometry and use the formalism obtained to design an experiment to detect source noise with the MWA;
in \sect~\ref{sec:observations} we describe the observations and analyse the results;
in \sect~\ref{sec:discussion} we discuss the implications of our results.

\section{REVIEW OF SOURCE NOISE}\label{sec:review}
Below, we draw on the literature to describe the properties of source noise.
We refer the reader to other articles which discuss the effect of source noise in other specific cases such as pulsar scintillometry \citep{2013ApJ...768..170J}, spectral line observations \citep{2002PASP..114.1087L}, pulsar timing \citep{2011MNRAS.418.1258O}, polarimetry \citep{1994IAUS..158...95D} or spectropolarimetry \citep{evla:159}.
The effect of scattering on source noise is discussed in \citet{1987RaSc...22..469C}.

Measurements of the statistics of source noise on very short time scales have also been used to try to detect coherent emission from pulsars \citep[][and references therein]{2003A&A...405..795S}.
In the current work we focus on the typical synthesis imaging case where the a large number of Nyquist samples are averaged in the correlator, and the integration time is far longer than any timescale on which source noise could conceivably be correlated.
In this case the central limit theorem will ensure that source noise is highly Gaussian.
We return briefly to this topic in \sect~\ref{sec:conclusion}.

Both sources and noise can be described either with a characteristic temperature, or as a flux density.
We follow \citet{1989AJ.....98.1112K} and \citet{1993AJ....106..797M} and use the source flux density $S$ and Noise Equivalent Flux Density $N$ throughout.
The latter, more commonly known as System Equivalent Flux Density (SEFD) is the flux density of a source that would double the power received by a single element. 
Thus, at the output of a single element there is the sum of a power proportional to $S$ and power proportional to $N$, where the constant of proportionality is the product of the collecting area of the antenna, the bandwidth of the receiver, and the antenna gain (and various efficiency parameters).
Note that $N$ includes not only receiver noise but any noise other than $S$.
This is justified below.

Throughout, we assume that all interferometer elements are identical, and that both $N$ and $S$ are stationary, stochastic and ergodic and so $2 B \tau$ samples of baseband data are independent.
For an unpolarised source, the number of independent samples can be doubled by observing two orthogonal polarisations, so an implicit $n_\mathit{pol}$ can be assumed alongside $B$ and $\tau$ in the equations below.
Polarisation of any kind implies correlation of some kind between orthogonal polarisations \citep{1999ASPC..180..671R} which will complicate the picture.
We refer the reader to the references given above for more details on the source noise associated with polarised sources or in polarimetric images.

In the usual case that $S\ll N$, the RMS noise in a synthesis map is given by 
\begin{equation}
	\label{eqn:weak}
	\sigma = \frac{N}{\sqrt{2 n_b B \tau}}
\end{equation}
where $N$ is the SEFD in jansky, $n_b$ is the number of baselines given by $n(n-1)/2$ where $n$ is the number of antennas (we assume throughout that all cross correlations are used), $B$ is the bandwidth in Hz, and $\tau$ is the observing time in seconds.

A similar expression can be derived for $\sigma$ in the case that one of the antennas is used as a ``single dish''. 
In this case $\sigma$ will be $\sqrt{2}$ higher than for a single-baseline interferometer, since even a single baseline has two antennas and therefore two realisations of noise \citep[see][\eqn~7--34]{1989ASPC....6..139C}.
Equivalently, the noise on a single baseline will be shared randomly between the real and imaginary parts of the visibility, and on average this will reduce the amplitude by $\sqrt{2}$.

In order to understand the how the properties of source noise differ from this weak-signal case, it is instructive to examine it in the strong source limit (i.e. where other sources of noise are negligible).
\citet{1989ASPC....6..431A} show that in this case the RMS map noise at the location of the source is simply
\begin{equation}\label{eqn:strong}
	\sigma = \frac{S}{\sqrt{B \tau}} ,
\end{equation}
replicating the single-dish result.
This is intuitive since all antennas see the same noise, and so unlike the weak-source case, the noise cannot be reduced by averaging over baselines.

\citet{1989ASPC....6..431A} derive the properties of source noise in various other regimes to draw a number of important qualitative conclusions:

Firstly, if a strong source is fully resolved on all baselines, then it will be uncorrelated between antennas and so will behave like system noise.
This is relevant for low-frequency interferometers such as the MWA where $N$ is actually dominated by Galactic synchrotron emission, which is almost completely resolved on all but the shortest baselines, and totally absent from the longer $>100\lambda$ baselines typically used for continuum imaging.
This provides the justification for lumping sources such as synchrotron radiation from our Galaxy, the cosmic microwave background, atmospheric noise, etc. in with $N$; and provides a more precise definition of $S$ as noise which is correlated between antennas.

Secondly, the noise only appears in the image where the source does; therefore, for a point source, the source noise level in the map is shaped like the instantaneous PSF of the array.  

Finally, they note that for an interferometer with $n$ elements, the source noise in the image will be comparable with the system noise when $N\sim n S$.
In \tab~\ref{tab:nn} we list $N/n$ for a number of interferometers to emphasise that this ``moderately strong source regime'' can easily be reached.
\begin{table}
  \footnotesize
  \centering
  \caption{\label{tab:nn}$N/n$ for various radio interferometers, built and planned. All come from the SKA Baseline Design\protect\footnotemark\ \tab~1 except the MWA figure which is derived from \citet{2013PASA...30....7T}.}
  \begin{tabular}{lr}
    \hline
    Instrument & $N/n$ \\
               &  Jy   \\
    \hline                                                                                                                     
    MWA        &  400 \\
    LOFAR      &  45.2 \\
    ASKAP      &  42.5 \\
    Meerkat    &  8.6 \\
    SKA-low    &  2.8 \\
    SKA-mid    &  1.7 \\
    \hline
  \end{tabular}
\end{table}
\footnotetext{SKA-TEL-SKO-DD-001 \url{http://www.skatelescope.org/wp-content/uploads/2012/07/SKA-TEL-SKO-DD-001-1_BaselineDesign1.pdf}}

\citet{1991ASPC...19....6A} derive a very simple expression for the noise (system noise and self noise) at any location in the map for a total power image.
We modify this slightly so that it simplifies to \eqns~\ref{eqn:weak}\&\ref{eqn:strong} in the appropriate limits:
\begin{equation}\label{eqn:weakstrong}
	\sigma =    
	\frac{1}{\sqrt{B\tau}}\left(S + \frac{N}{\sqrt{2 n_b}}\right)
\end{equation}
where $S$ is the apparent source brightness (source brightness distribution convolved with the PSF) at any point in the image.

This expression is only approximate for an interferometry image because it does not account precisely for how the noise on different baselines are correlated in this ``moderately strong source'' regime. 
A more complete treatment was given by \citet{1989AJ.....98.1112K}, who considered the correlation between each pair of \emph{baselines}.
Three classes of baseline pairs were identified: the autocorrelations of each baseline, those where the pair of baselines have an element in common, and those where the pair of baselines is composed of 4 different antennas.
Source noise correlates between the baselines differently in each case.
For a point source at the phase centre, the visibility on each baseline is identical, and the on-source noise is
\begin{equation} \label{eqn:m93}
	\sigma =
	\frac{S+N}{\sqrt{2 n_b B \tau}}\sqrt{1 + 2\left(n\!-\!2\right)S' + \left[1 + \left(n\!-\!1\right)\left(n\!-\!2\right)\right]S'^2}
\end{equation}
where
\begin{equation} \label{eqn:sprime}
	S' = \frac{S}{S+N} .
\end{equation}
Here we use the form of the equation given by \citet{1993AJ....106..797M} who corrected a minor error in \citet{1989AJ.....98.1112K}.
The three terms reflect the three classes of baseline pair: the autocorrelations will correlate perfectly regardless of whether $S$ or $N$ dominates and the other two classes scale with $S'$ and $S'^2$ respectively.

\Eqn~\ref{eqn:m93} reduces to \eqn~\ref{eqn:weak}\ or \ref{eqn:strong} in the appropriate limits.
It is also well-approximated by \eqn~\ref{eqn:weakstrong} for large $n$ as predicted by \citet{1991ASPC...19....6A} (the fractional error is $<1\%$ for $n>12$).

\begin{figure} 
  \includegraphics[width=\columnwidth]{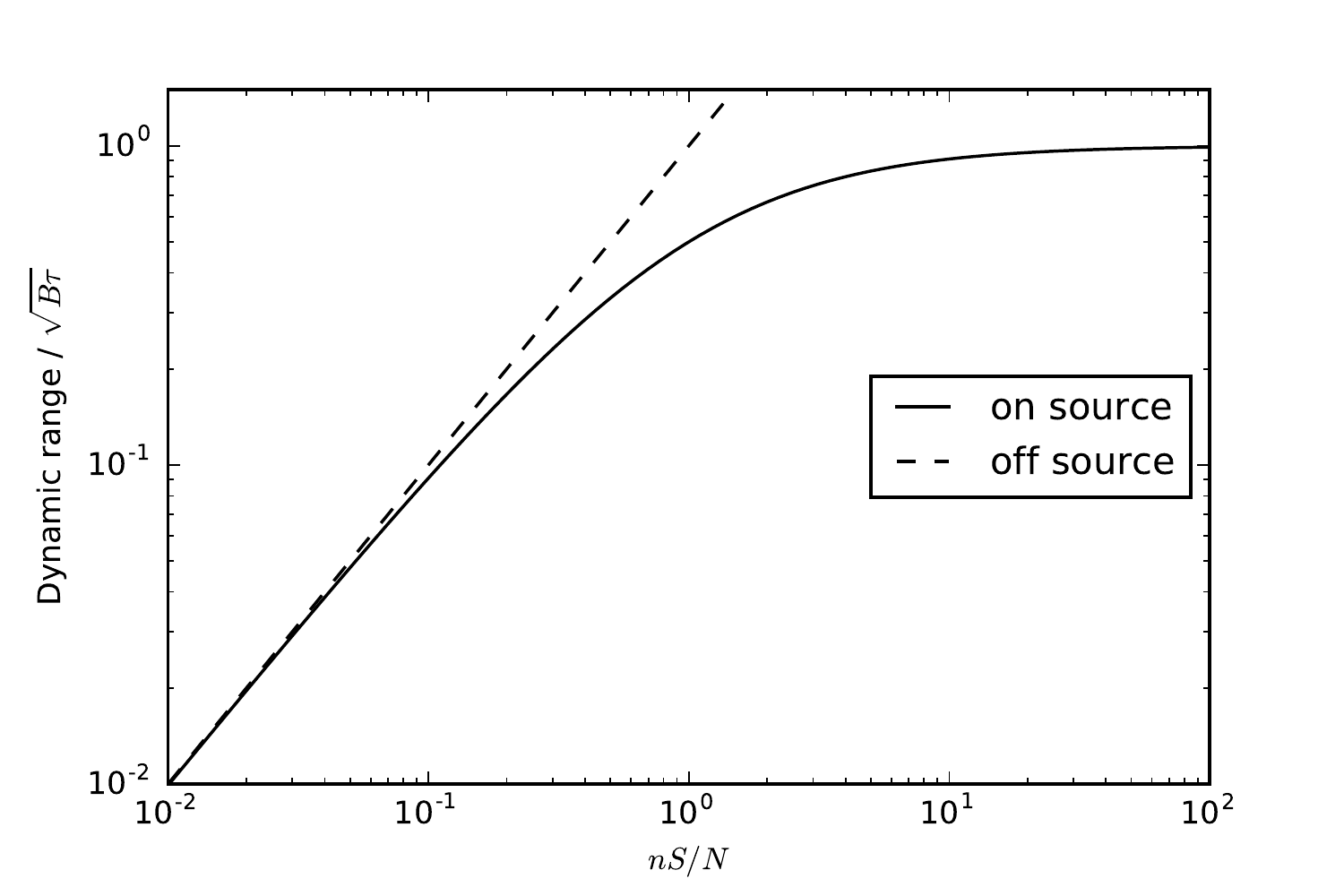}\,
  \caption{\label{fig:dynamicrange}Following \citet{1989AJ.....98.1112K} \fig~2, Maximum achievable dynamic range achievable off-source (dashed line; \eqn~\ref{eqn:weak}) and on-source (solid line; \eqn~\ref{eqn:weakstrong}) for as a function of source strength for a large-$n$ interferometer. This is generalised by measuring source flux density in units of $N/n$ and dynamic range as a fraction of its maximum ($\sqrt{B\tau}$).}
\end{figure}
\Fig~\ref{fig:dynamicrange} shows the achievable dynamic range (signal-to-noise) off- and on-source (i.e. calculated using \eqns~\ref{eqn:weak}\&\ref{eqn:m93} respectively). 
The source brightness is given in units of $N/n$ (see \tabs~\ref{tab:nn}).
On- and off-source noise start to diverge at $\sim N/10n$ -- just a few hundred mJy for the SKA; and the on-source dynamic range has almost reached its maximum value by $\sim10N/n$.

\Tab~\ref{tab:bt} shows the maximum dynamic range obtainable for various common MWA observing parameters.
For a typical snapshot, the dynamic range will be  $>10^4$ before source noise effects become noticeable.
This level of source noise is not easily measured since noise due to confusion noise or calibration errors are likely to limit the dynamic range far more.

In contrast, if an image is made with the smallest possible bandwidth and observing time with the current MWA correlator, the maximum achievable dynamic range is $\sim$100.
Solar images are regularly made in this way with the MWA, and the quiet Sun flux density at 150\,MHz is sufficiently far into the strong source regime that even if this flux density is spread over many pixels the brightness of each resolution unit to be close to the strong source regime.
This means that every resolution unit of the Sun will fluctuate by $\sim$1\% across timesteps, spectral channels and polarisations regardless of any intrinsic change in brightness.
However, rapid intrinsic changes in solar emission (which may be polarised) in both time and frequency would complicate measurements of source noise in observations of the Sun.
\begin{table}[t]
  \footnotesize
  \centering
  \caption{\label{tab:bt} $\sqrt{B\tau}$ for various MWA observing modes. All bandwidths take into account discarded band edges where appropriate. Maximum resolutions in time and frequency refer to the original online MWA correlator (still standard at the time of writing).}
  \begin{tabular}{rrrl}
\hline
$B$	&$\tau$	&$\sqrt{B\tau}$	& Use \\
MHz 	&  s	& 		&     \\
\hline                                                                                                                     
26.88	& 120.0	& 56794		& typical snapshot observation \\
13.44	& 0.5	& 2592		& IPS observing parameters (see \S~\ref{sec:observations}) \\
0.01	& 120.0	& 1095		& maximum spectral resolution  \\
0.04	&  0.5	& 141		& minimum time--bandwidth product \\
\hline
  \end{tabular}
\end{table}

Both spectral line and 0.5\,s snapshot imaging will start to show the effects of source noise at dynamic ranges $\sim$1000.
Thus if images could be made for each of a large number of spectral channels or timesteps the RMS of pixels on and off source would be expected to show a measurable difference.
The former is the approach taken by \citet{1993AJ....106..797M} in measuring source noise in the strong regime.
We take the latter approach which has much in common with the procedure used for making IPS observations with the MWA \citep{2018MNRAS.473.2965M}.

\section{OBSERVATIONS}\label{sec:observations}
As part of our Interplanetary Scintillation (IPS) observing campaign \citep{2019PASA...36....2M} we observed a series of calibrators in high time resolution mode with 5-minute observing time.
These observations pre-date the upgrade of the MWA to Phase II \citep{2018PASA...35...33W}.
Virgo A (M87), at our observing frequency of 162\,MHz was observed at 2016-01-16T21:04:55 UTC when it was 53 degrees above the horizon and 112 degrees from the Sun.
The solar elongation is relevant since Virgo A has a compact core and jet several janskys in brightness at GHz frequencies \citep[e.g.][]{1982ApJ...263..615R} which would be compact on IPS scales.
This means that Virgo A will also show interplanetary scintillation.
However this variability should be fully resolved with 2\,Hz sampling.
Throughout, we assume a flux density of 1032\,Jy for Virgo A \citep[based on values gleaned from the literature by ][]{2017MNRAS.464.1146H}.
We do not include any uncertainty on the flux density of Virgo A in our errors below (such an error would cause an equal scaling error in all of our measured and derived quantities).
Note that no direction-dependent flux scaling was carried out.

\begin{figure*} 
  \centering
  \includegraphics[width=0.75\textwidth]{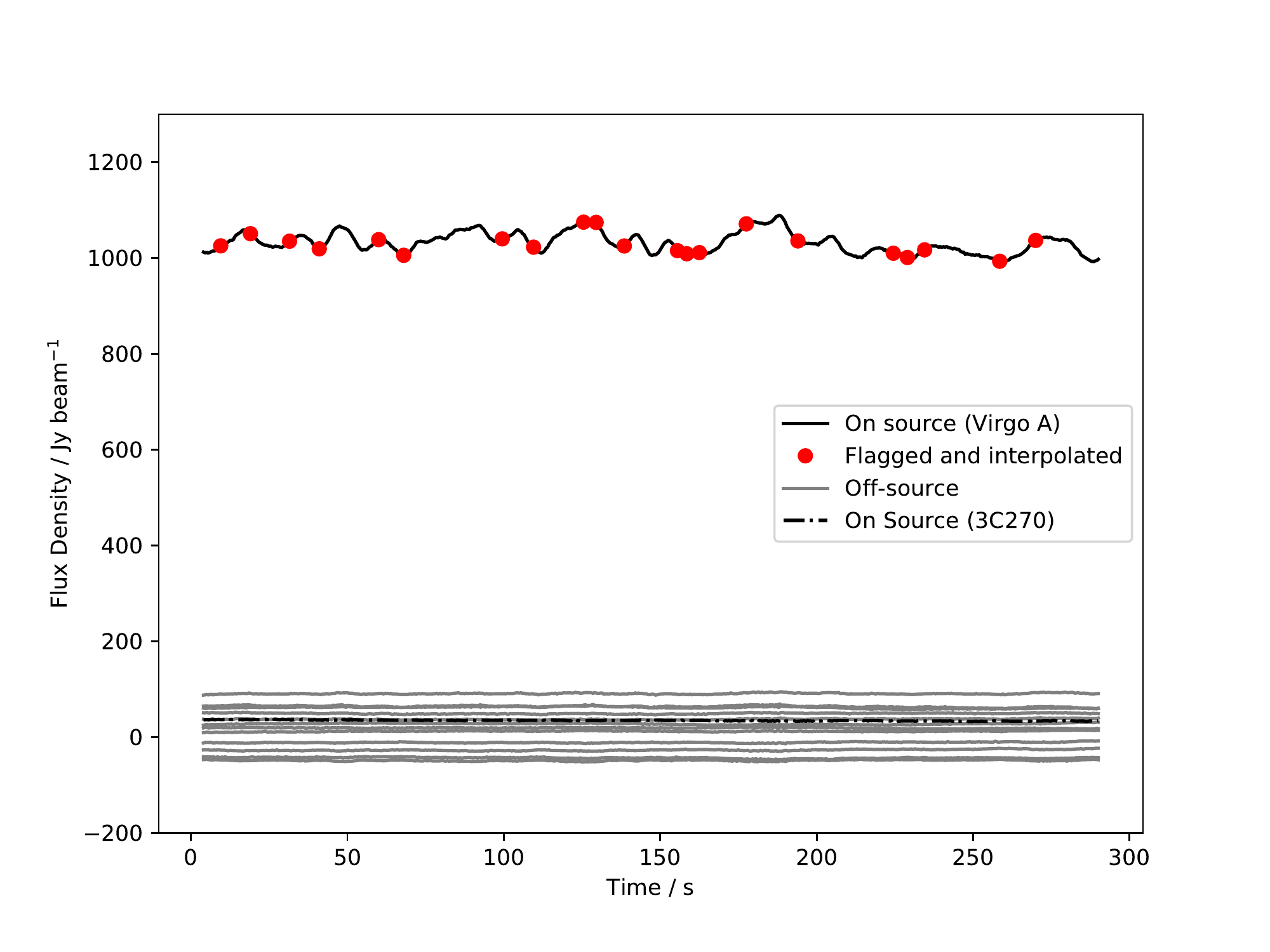}\,
  \caption{\label{fig:timeseries} Timeseries (lightcurve) showing brightness of various pixels in the image as a function of time. Black line shows the on-source lightcurve; dot-dashed black line shows a much weaker source: the (Western lobe of) 3C270; grey lines show a selection of off-source pixels. 21 points in the lightcurve had to be flagged due to clearly discrepant points, and these have been linearly interpolated over in all lightcurves; red circles denote these flagged points for the on-source pixel.}
\end{figure*}
Each individual 0.5\,s integration of the full observation was imaged separately for each instrumental polarisation (XX\&YY) using normal weighting and excluding baselines shorter than 24-$\lambda$ (a negligible fraction of the total number of baselines).
Throughout this paper we only refer to results derived from the XX images, however using the YY images produces qualitatively identical results.
\fig~\ref{fig:timeseries} shows the lightcurves for selected pixels in the image: the pixel corresponding to the brightest point in Virgo A and several pixels distributed through the image a few resolution units from Virgo A.
Additionally we show a pixel corresponding to the brightest point in 3C270, a resolved double with a peak apparent brightness approximately 3\% of that of Virgo A.
After standard flagging of the start and end of the observation, 573 timesteps remained.
A number of clearly discrepant points were easily discernible in all lightcurves (the same in each) and these have been interpolated over for all lightcurves.
The on-source lightcurve shows clear variability on a timescale consistent with ionospheric scintillation.

\begin{figure*} 
  \centering
  \includegraphics[width=0.75\textwidth]{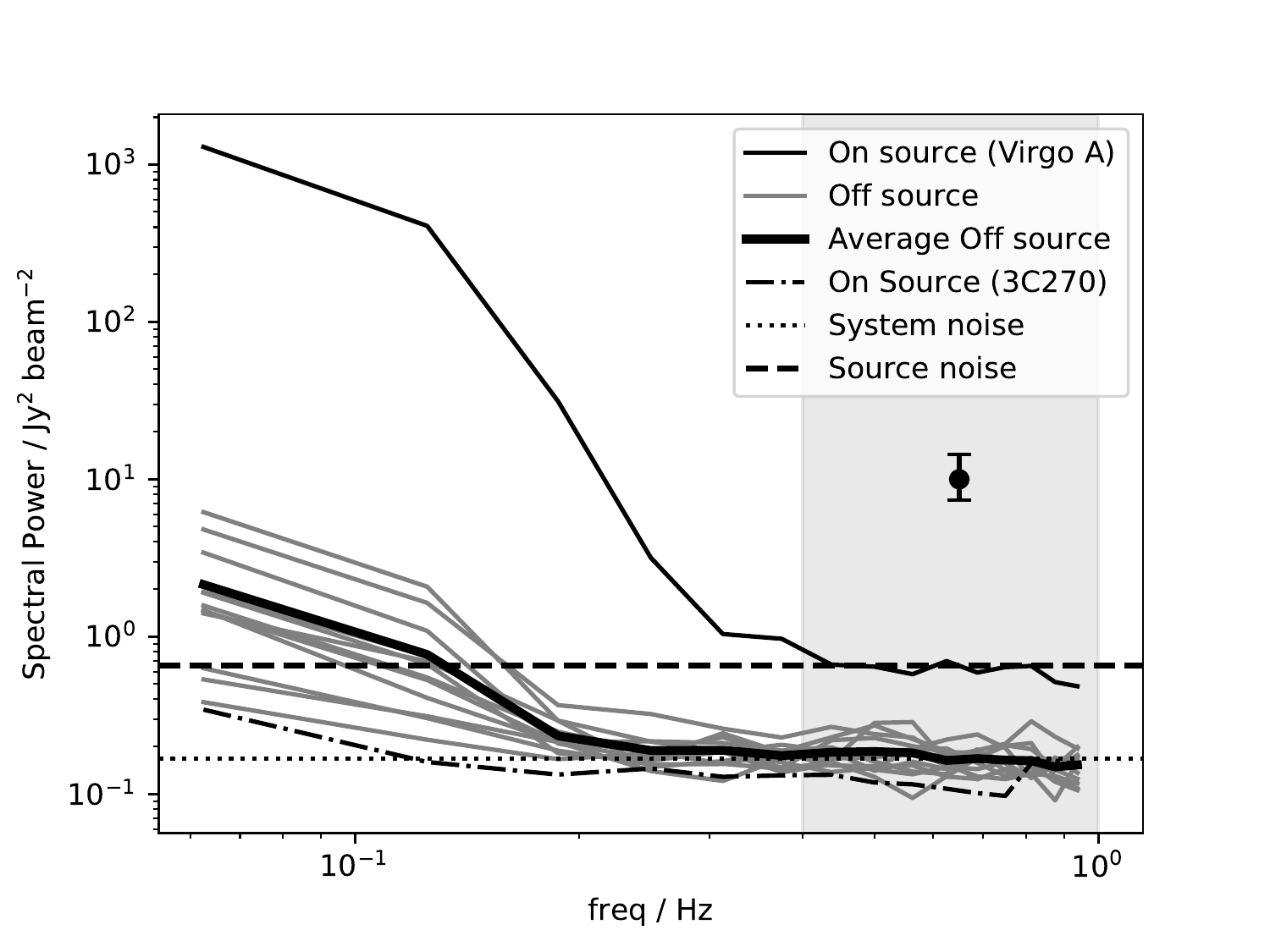}\,
  \caption{\label{fig:powerspectrum}Power spectrum for each of the lightcurves described in \fig~\ref{fig:timeseries}. The thin black line shows on-source power spectrum; the dot-dashed line shows a much weaker source (3C270); the grey lines show off-source power spectra, with the thick black line showing the average of the grey lines. Power spectrum parameters are described in the text. The single error bar shows the 95\% confidence interval for a single point with these parameters. The dotted line is the mean of all off-source power spectrum points above 0.4Hz. The dashed line is the estimate of the on-source power based on the on-source brightness and off-source noise.}
\end{figure*}
\fig~\ref{fig:powerspectrum} shows power spectra for each of the lightcurves shown in \fig~\ref{fig:timeseries}.
These were generated using Welch's method \citep{welch:1967} with 17 overlapping FFTs of 32 samples with a Hanning window function.
The strong signal that dominates at low frequencies is consistent with ionospheric scintillation and shows the characteristic Fresnel ``Knee'' at about 0.1\,Hz.
Above this frequency the power drops steeply.
The same ionospheric scintillation can be seen for all off-source lightcurves due to the sidelobes of Virgo A, however the sidelobes level is low enough that the variance is suppressed by at least 3 orders of magnitude.
3C270 scintillates independently of Virgo A and also has a slightly lower noise level\footnote{There is no obvious reason for the lower noise level in the vicinity of 3C270, and we believe it is due to some combination of factors, none of which are consequential enough to affect our analysis or conclusions.}.

Above 0.4\,Hz the variance from ionospheric scintillation is negligible, and the power spectrum consists only of white noise.
We conclude that all variability due to calibration errors such as those introduced by the ionosphere are restricted to lower frequencies as we would expect.
This white noise is clearly at very different levels on-source and off-source.
We can use the average of all points above 0.4\,Hz to measure the off-source noise to be 0.410$\pm$0.005\,Jy, and the on-source noise to be 0.75$\pm$0.03\,Jy.
The former implies that the SEFD ($N$) is 125000$\pm$1500\,Jy ($B=13.44$\,MHz, $\tau=0.5$\,s, $n=118$ i.e. 10 antennas flagged).
This is about a factor of 2.4 higher than that given in \tab~\ref{tab:nn}, which is not surprising given the off-zenith pointing and the fact that \citet{2013PASA...30....7T} assume a nominal sky temperature and field of view.

Using the source brightness and off-source noise we can calculate what the on-source noise should be using \eqn~\ref{eqn:m93}.
This prediction (0.807$\pm$0.005\,Jy) is shown as the dashed line on \fig~\ref{fig:powerspectrum} and it agrees extremely well with the measured on-source noise.
There is a small discrepancy, which is within 2-$\sigma$; however even this small excess can be explained as a small leakage of source noise into the off-source pixels due to the sidelobes of Virgo A.
Alternatively, this may be due to Virgo A being slightly resolved on some MWA baselines.

This does not leave any variance due to IPS.
IPS is extremely variable on the night-side, and the scintillation index would only be a few percent.
Furthermore, most of the variability is on timescales longer than 0.4\,Hz.
IPS may be responsible for the slight increase in power around 0.3\,Hz.

3C270 does not exhibit measurable source noise.
This is expected since, like the sidelobes of Virgo A, its brightness puts it well within the weak regime. 

\section{DISCUSSION}\label{sec:discussion}
We have detected source noise at a level which is firmly in line with the predictions of \eqn~\ref{eqn:weakstrong} (no empirical test for the existence of the subtle effects that differentiate \eqns~\ref{eqn:weakstrong}\&\ref{eqn:m93} are possible with our data since the two are practically identical for $n>100$).
We now consider the effect that this would have off-source in synthesis images. 

\subsection{The leakage of source noise off source}
\begin{figure} 
  \includegraphics[width=\columnwidth]{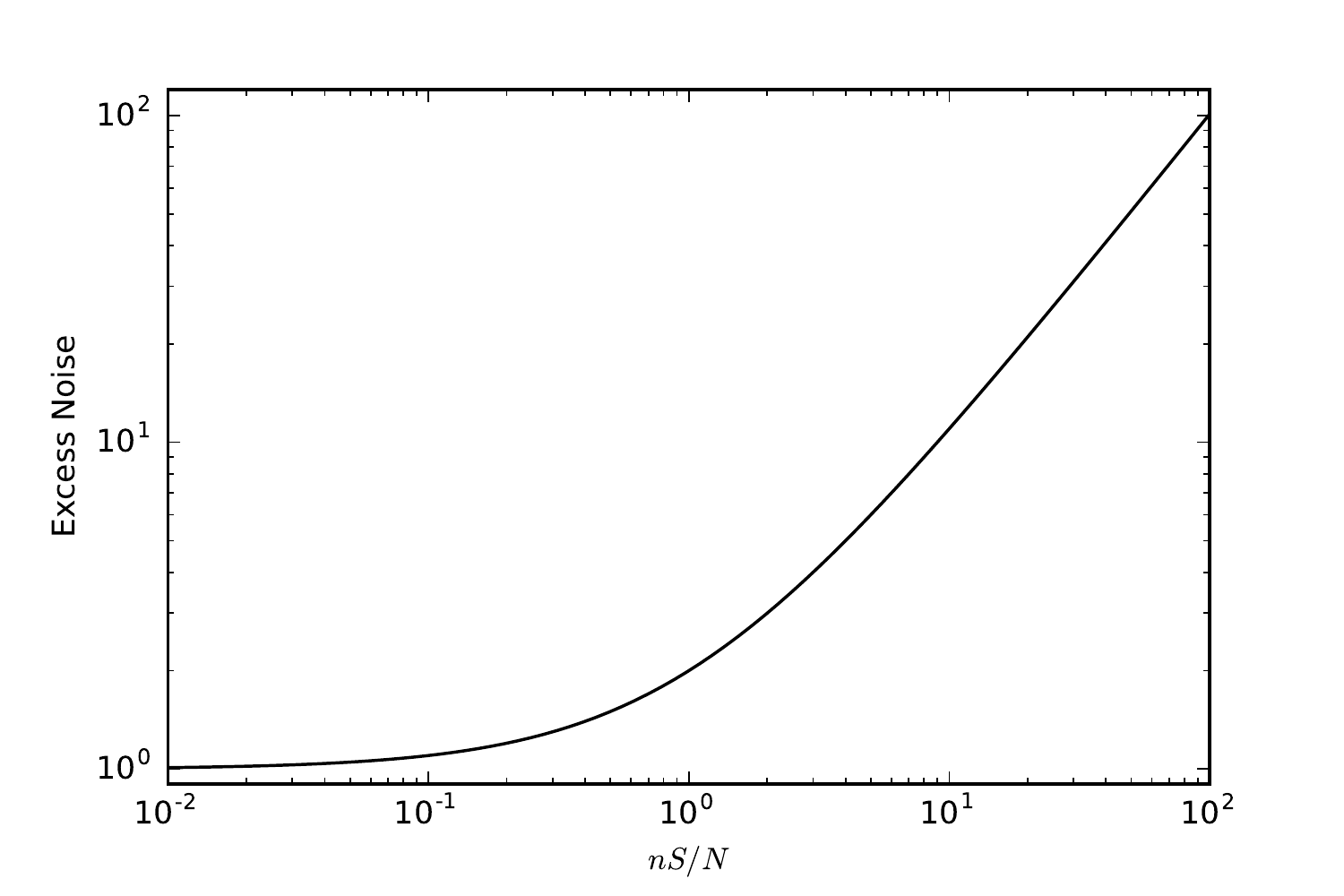}\,
  \caption{\label{fig:excess_noise} Ratio of on-source to system noise (\eqns~\ref{eqn:weakstrong}\&\ref{eqn:weak}) as a function of source strength for a large-N interferometer (this is the ratio of the dashed line to the solid line in \fig~\ref{fig:dynamicrange}).}
\end{figure}
\Fig~\ref{fig:excess_noise} plots the ratio of on-source noise to system noise where these are calculated using \eqns~\ref{eqn:weakstrong}\&\ref{eqn:weak} respectively.
To understand the effect this has off-source, consider a narrowband snapshot image.
In this case, the source noise will have the shape of the \emph{snapshot, monochromatic} PSF.
The sidelobes of this PSF can be characterised by their RMS relative to the central peak; for example, the proposed SKA-low configuration\footnote{SKA-SCI-LOW-001 \url{http://indico.skatelescope.org/event/384/attachments/3008/3961/SKA1_Low_Configuration_V4a.pdf}} has a sidelobe level of 1\% in the instantaneous monochromatic case.
Therefore for $S=N/n=2.7$\,Jy, the source noise level on-source will be $\sim2\times$ the weak-noise level, and off-source this 100\% increase in noise will become just 1\%.

As the bandwidth and integration time are increased, both the system noise and the source noise will reduce with $\sqrt{B\tau}$; the ratio between the two will remain constant, and the excess noise due to the source noise will remain at 1\%.
With appreciable fractional bandwidth and/or Earth rotation synthesis this source noise will be spread smoothly over the image.

For the example given, the effects of source noise off-source are extremely low; however there will be 5$\times$ 2.7\,Jy sources per FoV with the SKA-low \citep{2016MNRAS.459.3314F}.
For a 14\,Jy source ($\sim$1 per FoV) there would be an excess of 5\%. 
Only for a 270\,Jy source will the noise be doubled and such sources are relatively rare for extragalactic fields assuming arcminute or better resolution.

Two effects will reduce the impact of source noise even further.
First, the 1\% level given for the SKA-low case is for the 100 resolution units closest to the source.
The sidelobe level will drop even lower with increasing distance from the bright source.
Secondly, this level of source noise assumes natural weighting of baselines.
More uniform weighting will increase the weighting of certain baselines (the longer ones) by a large amount compared to the shorter spacings, particularly for instruments like the SKA-low with very high concentrations of collecting area at the centre.
This will reduce the effective number of antennas and therefore the source noise relative to the system noise.

\subsection{Subtracting Source Noise}
\citet{1989ASPC....6..431A} note that source noise can be subtracted perfectly from a snapshot by deconvolution, while noting the computational effort required to do so.
Here we explore the limitations on how cleanly this can be done.

Subtracting source noise requires that the field be imaged with sufficient time and frequency resolution for changes in the PSF to be insignificant from one image to the next.
The required resolution will depend on the array and imaging parameters, however the requirements are similar to those required to minimise time-average and bandwidth smearing \citep[e.g.][6.3-6.4]{2017isra.book.....T} and are likely to be demanding.
Here we concentrate on a more fundamental issue that such a subtraction poses: namely that if the source noise is to be perfectly subtracted, the source brightness must be allowed to vary as a function of time and frequency and no spectral smoothness can be assumed.
Conversely if spectral smoothness is strictly imposed, as is normally the case when subtracting continuum from spectral line cubes, no source noise will be subtracted.
Clearly any number of compromise schemes between these to extremes could be devised; however the principle remains that source noise can only be subtracted to the extent that it can be separated from any spectral or temporal signal that needs to be preserved.

\subsection{Conclusion}\label{sec:conclusion}
The properties of source noise lead us to the striking conclusions that the SKA will not more accurately characterise a 1000\,Jy source than a 10\,Jy source, and that the standard equations underestimate the noise on the measurement of a  2\,Jy source by a factor of two.
We have demonstrated that even the MWA---which is orders of magnitude less sensitive than the SKA---can measure these effects at the 12-$\sigma$ level (albeit it in an observation of a 1000\,Jy source contrived for the purpose).

Our observations show that the magnitude of the effect is precisely in line with predictions, and so source noise can be predicted very precisely from easily measurable parameters (the ratio of source brightness to system noise).
This is in contrast to other sources of variability such as the ionosphere.
Source noise is also expected to have stochastic behaviour as a function of frequency and time, and we exploit this to separate source noise from much stronger ionospheric effects.

Although the effects on standard continuum imaging appear to be benign in all but the most extreme cases, source noise could become problematic wherever weak signals \citep[including polarised signals, see][]{evla:159} need to be measured in the presence of strong emission; especially where the signals involved have fine frequency structure or vary on short timescales.
This situation will occur for any spectral line observations where the continuum has to be subtracted to see the much weaker line emission.
In this situation the noise will be higher and not uniform across the image.
Noise estimates made from regions of the image with no continuum will underestimate the noise.

Fast transients such as Fast Radio Bursts (FRBs) and pulsars will also reach the strong source limit, especially since their signals occupy only a narrow sloping band in a dynamic spectrum.
For ASKAP, a few of the detected FRBs \citep{2018Natur.562..386S} already have peak flux density in the strong source limit and for SKA sensitivity many will be in this regime.
Thus measurements of source noise of FRBs should be possible, and such measurements may be valuable, since they probe the extent to which the emission is coherent \citep{2009IAUS..257..305M}.

\section*{Acknowledgements}
We would like to thank J-P Macquart, Adrian Sutinjo and Wasim Raja for useful discussions.
This scientific work makes use of the Murchison Radio-astronomy Observatory, operated by CSIRO.
We acknowledge the Wajarri Yamatji people as the traditional owners of the Observatory site.
Support for the operation of the MWA is provided by the Australian Government (NCRIS), under a contract to Curtin University administered by Astronomy Australia Limited.
We acknowledge the Pawsey Supercomputing Centre which is supported by the Western Australian and Australian Governments.

\bibliographystyle{pasa-mnras}
\bibliography{refs}
\end{document}